\documentclass[aps,twocolumn,groupedaddress,showpacs]{revtex4}

\usepackage{graphicx}
\usepackage{amsfonts}
\usepackage{amsmath}
\usepackage{amssymb}
\usepackage{bm}

\begin{document}

\title{Analytical theory and possible detection of the $ac$ quantum spin Hall effect}
\author{W. Y. Deng$^1$}
\author{Y. J. Ren$^1$}
\author{Z. X. Lin$^1$}
\author{L. Sheng$^{1,2}$}
\email{shengli@nju.edu.cn}
\author{D. N. Sheng$^3$}
\author{D. Y. Xing$^{1,2}$}
\affiliation{$^1$ National Laboratory of Solid State Microstructures and Department of Physics, Nanjing University, Nanjing 210093, China\\
 $^2$ Collaborative Innovation Center of Advanced Microstructures,
Nanjing University, Nanjing 210093, China\\
$^3$ Department of Physics and Astronomy, California State
University, Northridge, California 91330, USA}
\date{\today }

\begin{abstract}
We develop an analytical theory of the low-frequency $ac$ quantum spin Hall (QSH) effect
based upon the scattering matrix formalism.
It is shown that the $ac$ QSH effect can be interpreted as a bulk quantum pumping effect.
When the electron spin is conserved, the integer-quantized
$ac$ spin Hall conductivity can be linked to the winding
numbers of the reflection matrices in the electrodes, which also equal to the bulk
spin Chern numbers of the QSH material. Furthermore, a possible experimental
scheme by using ferromagnetic metals as electrodes is proposed to detect
the topological $ac$ spin current by electrical means.
\end{abstract}

\maketitle

\section{Introduction}\label{s1}

Topological insulators (TIs) are currently on the research front of condensed matter physics, because of their fundamental interest and potential applications in spintronic devices~\cite{qshe1,qshe2,qshe3,3dti1,3dti2,3dti3,rev1,rev2,rev3,rev4,rev5,Weyl,rev6}. Two-dimensional (2D) TIs are also called the quantum spin Hall (QSH) systems, as they can host
the interesting QSH effect, in which quantized spin current or spin accumulation
can be generated in response to an applied electric field. A QSH system is an insulator in the bulk with a pair of conducting gapless edge states traversing the bulk band gap~\cite{qshe1,qshe2,qshe3}. The existence of the edge states has its origin in the nontrivial bulk topological properties, which are usually characterized by the $Z_2$ index~\cite{Z2} or spin Chern numbers~\cite{spinch1,spinch2,spinch3}, but their gapless nature is protected by the time-reversal (TR) symmetry.
While the nondissipative electronic transport through the edge modes is immune to nonmagnetic disorder, the edge states will become gapped and Anderson localized in the presence of TR-symmetry-breaking perturbations~\cite{qshe1,Yang}. As a consequence, the QSH effect is often unstable in realistic environments. Up to now,
conductance through edge channels near the theoretically
predicted quantized value has been detected only in small samples of
HgTe quantum wells~\cite{HgTe} and InAs/GaSb
bilayers~\cite{RRDu}. Realizing
QSH effect that is as robust as the conventional quantum Hall effect is still challenging.

Recently, it was numerically demonstrated that the $ac$ QSH effect driven by an $ac$ electric field
is fundamentally different from the intensively researched $dc$ QSH effect~\cite{AcQSH}.
In contrast to the $dc$ QSH effect, which must be carried by edge states,
the $ac$ QSH effect can occur in the bulk without involving the fragile edge states,
hence being robust against TR-symmetry breaking and disorder.
In fact, $ac$ spin-dependent electronic transport has started to capture attention
in recent years, stimulating the emerging field of $ac$ spintronics. Jiao and Bauer theoretically predicted
that the much larger $ac$ voltage could be used to detect spin currents in the spin
pumping transport~\cite{Ac1}. Wei $et$ $al.$ found experimentally that the $ac$ spin
current is much larger than the $dc$ component in a ferromagnet-normal junction with
time-dependent magnetization vector~\cite{Ac2}.

In this paper, we show that the basic characteristics of the low-frequency $ac$ QSH effect can
be described by a simple analytical theory based on
the time-dependent scattering matrix formalism.
The $ac$ spin Hall conductivity  in the adiabatic regime
is linked to the winding numbers of the reflection matrices
in the electrodes, which also equal to the spin Chern numbers of the QSH material.
The theoretical result is consistent with the numerical calculation based upon
the Kubo linear-response theory~\cite{AcQSH}.
Our theory indicates that while the $ac$ and $dc$ QSH effects behave quite differently,
they share the same topological origin.
 We further show that when ferromagnetic metals
are used as electrodes, the topological $ac$ spin current will induce an
electrical voltage difference along the
electrodes, suggesting a possible experimental way to observe the $ac$ QSH effect
by electrical means.

In the next section, we present a general
theoretical description of the $ac$ QSH effect based on the
scattering matrix formula. In Sec.\ III, we consider the BHZ model as a concrete
example, and evaluate the
the winding numbers and spin Hall conductivity. In Sec.\ IV, a possible way to measure the $ac$ spin Hall conductivity experimentally is proposed. The final section contains a summary.

\section{A General Description}\label{s2}

Let us consider the setup illustrated in Fig.\ 1. Two metallic bars (blue) are deposited on
top of a QSH sample. When an $ac$ electric voltage difference is applied to the bars,
an $ac$ electric field $E(t)=E_{0}\cos(\omega t)$
is established between the bars, and in response, a Hall spin current $j_{s}(t)$
is generated in the biased region. Two metallic gates (red), $A$ and $B$, are deposited
between the bars near the edges of the biased region, which serve as source and
drain electrodes for the Hall spin current.
To facilitate our general discussion,
we assume that the $ac$ electric field
also exists in the electrodes. This assumption does not change
the topological properties of the system, and will not
affect the main conclusion. The sizes of the biased region and electrodes
are taken to be sufficiently large, so that we can neglect any finite-size effects.
As a result, the Hamiltonian of the
system of the biased region and electrodes
takes the form $H(k_{x},\tilde{k}_{y},x)$ with
$\tilde{k}_{y}=k_{y}-eA(t)$. Here, $(k_{x},k_{y})$ is the 2D momentum,
where $k_{x}=\frac{\hbar}{i}\frac{\partial}{\partial x}$ is an operator,
and $k_{y}$ is a good quantum number.
$A(t)=-\frac{E_{0}}{\omega}\sin(\omega t)$
is the vector potential of the $ac$ electric field
$E(t)=E_{0}\cos(\omega t)$. In the biased region, $H(k_{x},\tilde{k}_{y},x)$
is the Hamiltonian of the QSH material with a band gap, while in the electrodes,
it represents the Hamiltonian of a metal. The remaining region of the QSH sample, which is
neither excited by the $ac$ electric field nor in contact with the
electrodes, will behave just like an ordinary insulator~\cite{AcQSH}, and so is not included into
the Hamiltonian.

\begin{figure}
\includegraphics[width=2.8in]{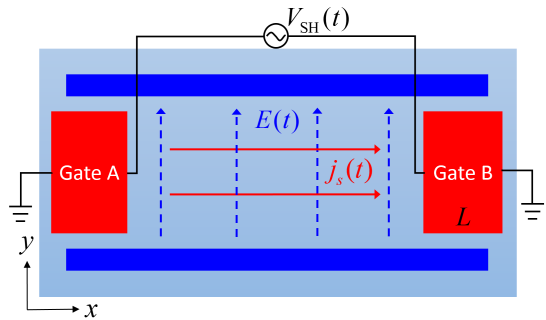}
\caption{A schematic view of a proposed setup to study the topological $ac$ QSH effect.
Two metal bars (blue) are deposited on top of a 2D QSH material.
When an $ac$ electrical voltage is applied to the metal bars, an electric field $E(t)$ will
be generated in the $y$ direction. In response, an $ac$ Hall spin current $j_{s}(t)$ is created along
the $x$ direction. Two metal gates (red) with length $L$, i.e., gates $A$ and $B$, are deposited
near the two edges of the biased region as source and drain electrodes for
the spin current. We first assume the electrodes to be nonmagnetic. Later,
we will show that if ferromagnetic metals are used as the electrodes,
an $ac$ electric voltage difference $V_{\mbox{\tiny SH}}(t)$ can be induced between
the inside edges of the two electrodes, suggesting a possible way to detect the $ac$ QSH
effect electrically.
}\label{Fig1}
\end{figure}

The electron Fermi energy is
set to be in the band gap of the QSH material.
We consider first the ideal case, where
the electron spin is conserved. We focus on the right electrode, and
the situation in the left electrode is similar. Let $\hat{r}_{ss}(\tilde{k}_{y})$
be the reflection matrix
for an electron with spin $s$ ($\uparrow$ or $\downarrow$) at the Fermi level
transmitting from the electrode toward the biased region.
In the adiabatic regime,
the spin current density pumped into
the electrode at time $t$ can be evaluated by using the
scattering matrix formula~\cite{Pump1,Pump2}
\begin{eqnarray}
j_{s}(t)=\frac{1}{L_{y}}\left(\frac{\hbar}{4\pi i}\right)\sum_{k_{y}\in\mbox{\tiny BZ}}
\mbox{Tr}\Bigl(\hat{r}^{\dagger}_{\uparrow\uparrow}\frac{d}{dt}\hat{r}_{\uparrow\uparrow}
-\hat{r}^{\dagger}_{\downarrow\downarrow}\frac{d}{dt}\hat{r}_{\downarrow\downarrow}\Bigr)\ ,
\label{spinJs1}
\end{eqnarray}
with $L_y$ as the cross-section of the biased region.
We notice that the reflection matrices depend on $t$ only through the variable
$\tilde{k}_{y}=k_{y}-eA(t)$, such that
$\frac{d}{dt} \hat{r}_{\uparrow\uparrow}=eE(t)\frac{d}{d k_{y}}\hat{r}_{\uparrow\uparrow}$, and $\frac{d}{dt}
\hat{r}_{\downarrow\downarrow}=eE(t)\frac{d}{d k_{y}}\hat{r}_{\downarrow\downarrow}$.
By using these relations and replacing
the summation over $k_{y}$ in Eq.\ (\ref{spinJs1}) by an integral, we derive
the $ac$ spin Hall conductivity, defined as
$\sigma_{\mbox{\tiny SH}}(\omega)=j_{s}(t)/E(t)$, to be
\begin{equation}
\sigma_{\mbox{\tiny SH}}(\omega)=\frac{e}{4\pi}(W_{\uparrow}-W_{\downarrow})\ ,
\end{equation}
where
\begin{equation}
W_{s}=\frac{1}{2\pi i}\int_{\mbox{\tiny BZ}}dk_{y}\mbox{Tr}
\left(\hat{r}^{\dagger}_{ss}\frac{d}{dk_{y}}
\hat{r}_{ss}\right)\ .
\label{Wind}
\end{equation}

Since the electron Fermi energy is in the band gap of the QSH material,
an electron incident from the electrode will be fully reflected, and
the reflection matrices must be unitary, i.e.,
$\hat{r}_{ss}^{\dagger}(\tilde{k}_{y})\hat{r}_{ss}(\tilde{k}_{y})=\hat{1}$.
Besides, they are periodic functions of $k_{y}$ in
a Brillouin zone. As a result, one can identify immediately $W_{s}$ as winding numbers,
which are always integers. Therefore, while the spin Hall conductivity $\sigma_{\mbox{\tiny SH}}(\omega)$
is defined as the ratio between two time-dependent quantities, i.e., the $ac$ spin current $j_{s}(t)$
and $ac$
electric field $E(t)$, it is integer-quantized
at any time, in units of the spin conductivity quantum $\frac{e}{4\pi}$,
in the adiabatic regime. Very often, continuous models are employed in
theoretical works. In a well-defined continuous model, the electron
wavefunctions should be continuous in the $k_{y}\rightarrow\pm\infty$ limit,
which implies
\begin{equation}
\lim_{k_{y}\rightarrow+\infty}\hat{r}_{ss}(\tilde{k}_{y})=\lim_{k_{y}\rightarrow-\infty}\hat{r}_{ss}(\tilde{k}_{y})\ .
\label{LimCond}
\end{equation}
Under this condition, $W_{s}$ remain to be integer-quantized. In the example considered later,
we will see that $W_{s}$ in fact equal to the spin Chern numbers of the QSH system.
When small spin-mixing perturbations, such as the Rashba spin-orbit coupling, are
present, spin-flip reflection processes will occur with small probabilities,
and the $ac$ spin Hall conductivity will deviate from the integer-quantized value
in a gradual manner, similarly to the $dc$ QSH effect.

We need to point out that the $ac$ Hall spin current originates from the time dependence
of the Hamiltonian caused by the $ac$ applied electric field, as
clearly indicated by Eq.\ (\ref{spinJs1}).
One might think that by taking the limit $\omega\rightarrow 0$,
the conclusion for the $ac$ QSH effect should be applicable to
the $dc$ QSH effect, which is not true.
For an exactly static electric field ($\omega=0$),
since one can choose an electrostatic scalar potential
to make the Hamiltonian independent of time, no spin current can
be generated in the setup shown in Fig.\ 1. The $dc$ QSH effect
must be carried by the edge states. Therefore, the $ac$ QSH effect
is substantially different from the $dc$ QSH effect. Moreover,
the above general discussion about the $ac$ QSH effect does not rely on any symmetries,
which is also different from the $dc$ QSH effect.
Small TR-symmetry-breaking perturbations will open
an energy gap in the spectrum of the edge states in a QSH material, and the edge states become
exponentially localized
immediately due to their one-dimensional nature. As a consequence, the $dc$ QSH
effect will be destroyed for sufficiently large samples. In contrast,
the $ac$ QSH effect is a bulk transport phenomenon, being
robust against TR-symmetry breaking and disorder~\cite{AcQSH}.

\section{A Concrete Example}\label{s3}

As a concrete example, we consider the BHZ model, which can be used
to describe the HgTe quantum wells~\cite{BHZ} or InAs/GaSb bilayers~\cite{InAs}.
The BHZ model Hamiltonian reads
\begin{equation}
H_{\mbox{\tiny QSH}}=v_{\mbox{\tiny F}}(k_{x}\hat{s}_{z}\hat{\sigma}_{x}-\tilde{k}_{y}\hat{\sigma}_{y})
-(M_{0}-B\tilde{k}^2)\hat{\sigma}_{z}\ .
\label{H_HgTe}
\end{equation}
Here, we retain the $B\tilde{k}^2$ term with $\tilde{k}^2=k_{x}^2+\tilde{k}_{y}^2$, as
it ensures that the condition Eq.\ (\ref{LimCond}) is fulfilled, and
the topological properties
of the system are properly defined. In fact, the spin Chern numbers of this
model, given by $C_{\uparrow(\downarrow)}=\pm\frac{1}{2}[\mbox{sgn}(M_{0})+
\mbox{sgn}(B)]$~\cite{HgTeSpinCh}, are dependent of $B$. Other nonessential nonlinear
terms of momentum in the original model have been neglected for simplicity.

The $ac$ QSH effect, as a topological transport phenomenon, is insensitive to the material
details of the electrodes. We model the electrodes by using a simple parabolic Hamiltonian
$H_{\mbox{\tiny E}}=-U_{0}+\frac{\tilde{k}^2}{2m}$. $U_{0}$ is taken to be large compared with
all other energy scales, so as to guarantee that the electrodes have sufficient number of conducting
channels for the spin current to flow through. Finite
potential barriers of height $V_{0}$ and thickness $d$ exist
at the interfaces between the QSH material and electrodes, similar to the
setup considered in Ref.~\cite{Cpump3}. By following the same procedure
detailed in Ref.~\cite{Cpump3}, linearizing the Hamiltonians of both the
QSH material and electrodes with respect to $k_{x}$, one can obtain for the
reflection coefficients
\begin{equation}
r_{\uparrow\uparrow}(\tilde{k}_{y})=-\frac{\cos(\theta)+i[\mbox{sh}(\gamma_{0}d)-\sin(\theta)
\mbox{ch}(\gamma_{0}d)]}{\mbox{ch}(\gamma_{0}d)-\sin(\theta)
\mbox{sh}(\gamma_{0}d)}\ ,
\label{Rupup0}
\end{equation}
and $r_{\downarrow\downarrow}(\tilde{k}_{y})=r_{\uparrow\uparrow}(\tilde{k}_{y})\vert_{\theta\rightarrow(\pi-\theta)}$,
where $\gamma_{0}=\frac{V_{0}}{\hbar}\sqrt{2m/U_{0}}$ and $\theta=
\mbox{arg}[v_{\mbox{\tiny F}}\tilde{k}_y+i(M_{0}-B\tilde{k}_{y}^2)]$.

For the present model, the reflection matrix $\hat{r}_{ss}(\tilde{k}_{y})$
is simply a number, satisfying $\left| {{r_{ss}}} \right|^2 = 1$. Therefore,
with changing $k_{y}$ from $-\infty$ to $\infty$,
$r_{ss}(\tilde{k}_{y})$ keeps traveling on the unit circle
around the origin on the complex plane, and forms a closed orbit
due to single-value condition Eq.\ (\ref{LimCond}).
The quantity $W_{s}$ defined in Eq.\ (\ref{Wind}) is the winding number
of the closed orbit around the origin. For convenience, the winding number
may also be expressed as
\begin{equation}
{W_s} = \frac{1}{{2\pi }}\left[ {{\varphi _s}\left( \infty  \right) - {\varphi _s}\left( { - \infty } \right)} \right]\ ,
\label{Wind2}
\end{equation}
where $\varphi_{s}(k_{y})$ is the argument of $r_{ss}(\tilde{k}_{y})$.
In the absence of the potential barrier, i.e., $\gamma_{0}d=0$, the reflection amplitude Eq.\ (\ref{Rupup0})
reduces to $r_{\uparrow\uparrow}(\tilde{k}_{y})=-e^{-i\theta}$.
The winding number can be determined by
tracking how the argument $\varphi_{s}(k_{y})$ of $r_{ss}(\tilde{k}_{y})$
evolves with changing $k_{y}$ from $-\infty$ to $\infty$. In Fig.\ 2(a), we plot
four different representative behaviors of $\varphi_{\uparrow}(k_{y})$. For simplicity,
we have chosen the unit set, where $v_{\mbox{\tiny F}}=\vert M_{0}\vert=1$. From Fig.\ 2(a), we see that
if $M_{0}>0$ and $B>0$, $\varphi_{\uparrow}(k_{y})$ increments $2\pi$ with varying
$k_{y}$ from $-\infty$ to $\infty$. If $M_{0}<0$ and $B<0$, $\varphi_{\uparrow}(k_{y})$ decrements
$2\pi$. In the other cases, where $M_{0}>0$ and $B<0$, or $M_{0}<0$ and $B>0$,
$\varphi_{\uparrow}(k_{y})$ does not change. The behaviors of $\varphi_{\downarrow}(k_{y})$
can be analyzed similarly. Consequently, we obtain from Eq.\ (\ref{Wind2}) the following
expression for the winding numbers
\begin{eqnarray}
W_{\uparrow(\downarrow)}=\pm\frac{1}{2}[\mbox{sgn}(M_{0})+
\mbox{sgn}(B)]\equiv
C_{\uparrow(\downarrow)}\ .
\label{BHZ_Wind}
\end{eqnarray}
Interestingly, the winding numbers are exactly equal to the spin Chern numbers of the BHZ model.
As expected, the $ac$ spin Hall conductivity is integer-quantized,
${\sigma _{\mbox{\tiny SH}}} = [\mbox{sgn}(M_{0})+
\mbox{sgn}(B)]\frac{e}{4\pi}\equiv(C_{\uparrow}-C_{\downarrow})
\frac{e}{4\pi}$, which is consistent
with the numerical result calculated from the Kubo theory at low frequencies~\cite{AcQSH}. This relation
indicates that while the $ac$ and $dc$ QSH effects behave quite differently,
they share the same topological origin.

\begin{figure}
\includegraphics[width=2.8in]{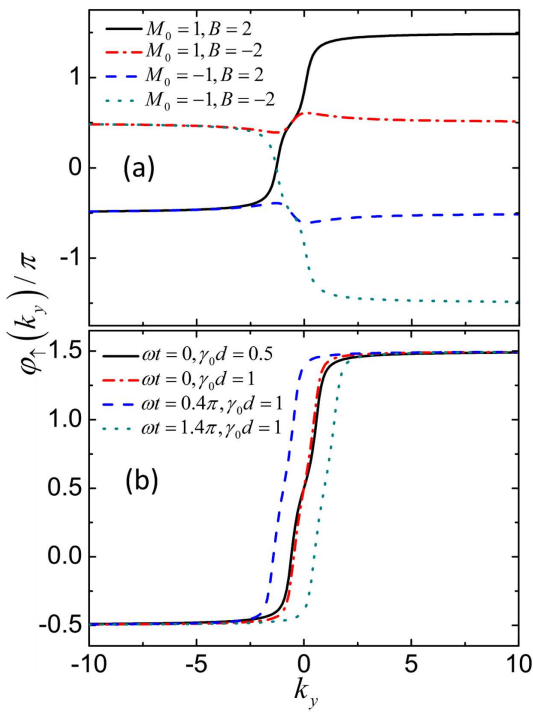}
\caption{Argument of the complex reflection amplitude, ${\varphi _ \uparrow }\left( {{k_y}} \right) = \arg \left( {{r_{ \uparrow  \uparrow }}} \right)$, as a function of $k_y$, for (a) four
different sets of $\left( {{M_0},B} \right)$,  and (b) four
sets of $\left( {\omega t,{\gamma _0}d} \right)$. The other parameters are taken to
be $e{E_0}/\omega  = 1$, and (a) ${\omega t = {\gamma _0}d = 0}$, (b) $M_{0}=1$, $B=2$.
}\label{Fig2}
\end{figure}

For a nonvanishing potential barrier, i.e., $\gamma_{0}d>0$, the condition $\left| {{r_{ss}}} \right|^2 = 1$
is still satisfied. This means that
with changing $k_{y}$ from $-\infty$ to $\infty$,
the reflection amplitudes $r_{ss}(\tilde{k}_{y})$ always move on the unit circle
around the origin on the complex plane. As a result, the winding numbers
cannot change values with changing $\gamma_{0}d$. In other words, Eq.\ (\ref{BHZ_Wind}) remains
valid for a nonvanishing potential barrier. Here, based upon the same topological argument,
we may also get some insight into why the $ac$ spin Hall conductivity
$\sigma_{\mbox{\tiny SH}}(\omega)=j_{s}(t)/E(t)$ is integer-quantized, being independent of
time. At a given time $t$, the system Hamiltonian $H(t)$ has some small deformation
from $H(t=0)$. As long as the difference $H(t)-H(0)$ is not large enough to close
the band gap, the winding numbers and spin Hall conductivity are unchangeable. In Fig.\ 2(b), we plot
the argument $\varphi_{\uparrow}(k_{y})$ of $r_{\uparrow
\uparrow}(\tilde{k}_{y})$ for some different sets of $\gamma_{0}d$ and $\omega t$
in the case $M_{0}>0$ and $B>0$. We see that while
its curve deforms with changing $\gamma_{0}d$ or $\omega t$, $\varphi_{\uparrow}(k_{y})$
always increments $2\pi$,
independent of the barrier strength or time. For essentially the same reason,
it is easy to understand that whether the $ac$ electric field exists in the electrode does not affect
the expressions for the winding numbers and $ac$ spin Hall conductivity.
The winding numbers may change values, only if the bulk band gap in the QSH material
closes. In this case, the condition $\left| {{r_{ss}}} \right|^2 = 1$ of full reflection no longer holds,
and the trajectories of $r_{ss}(\tilde{k}_y)$ can sweep
across the origin on the complex plane, which leads to a change in the winding numbers,
signaling a topological phase transition.

Similarly to Ref.\ ~\cite{Cpump3}, one can also include the
Rashba spin-orbit coupling into $H_{\mbox{\tiny QSH}}$,
and find that it leads to small deviation
of the spin Hall conductivity from the integer-quantized value, in the second order
of $R/v_{\mbox{\tiny F}}$ with $R$ as the strength of the
Rashba spin-orbit coupling.

\section{Experimental Measurement of The Ac QSH Effect}\label{s4}

Now we propose a method to experimentally measure the
$ac$ spin Hall conductivity  by electrical means. For the same setup shown in Fig.\ 1,
we suggest to use ferromagnetic metals as gates $A$ and $B$, with their magnetic moments
aligned along the $z$ axis.
The ferromagnets are assumed to have a length $L$, in the $x$ direction, much greater than the
spin diffusion length $\ell_{\mbox{\tiny D}}$.
As demonstrated above, the spin current generated is topological and insensitive
to the material parameters of the electrodes.
When the $ac$ electric field $E(t)$ is applied, a pure $ac$ spin current
density $\sigma_{\mbox{\tiny SH}}E(t)$ will flow
out of the source electrode and into the drain electrode.
The principle of the proposed method can be explained as follows.
The amplitude of the spin current decays into
the electrodes within about a spin diffusion length $\ell_{\mbox{\tiny D}}$,
and its spin-up and spin-down components recombine to cancel each other .
Since the majority-spin and minority-spin bands of a ferromagnet have different
2D conductivities, ${\sigma _ {\mbox{\tiny M}}}$ and ${\sigma _ {\mbox{\tiny m}}}$,
the voltage drops (strictly speaking, drops of the electrochemical potential)
in the two spin channels caused by the spin current
are not equal in magnitudes, giving rise to a net $ac$ electric voltage drop.
If the voltages at the outside edges
of the ferromagnetic electrodes are made equal through grounding
or short-circuiting, as shown in Fig.\ 1,
a voltage difference $V_{\mbox{\tiny SH}}(t)$ will appear between the inner edges of the
electrodes, which can be measured by using an $ac$ voltage meter.

To calculate the voltage difference, we employ the semiclassical spin diffusion equation for
a ferromagnetic metal, which can be written as~\cite{Diffusion}
\begin{equation}
{\nabla ^2}{\mu _s}\left( x,t \right) = \frac{{{\mu _s}\left( x,t \right) - {\mu _{ - s}}\left( x,t \right)}}{{l_s^2}}\ .
\label{DiffuseEq}
\end{equation}
Here, ${l_s } = {v_{\mbox{\tiny F}}}\sqrt {{\tau _s }{\tau _{ \uparrow  \downarrow }}/3} $, where ${{\tau _s }}$ and ${{\tau _{ \uparrow  \downarrow }}}$ are the electron non-spin-flip and spin-flip relaxation times,
${\mu _s }\left( x,t \right)$ is the spin-dependent electrochemical potential
for spin-up ($s =  \uparrow $) and spin-down ($s =  \downarrow $) electrons,
respectively, and ${v_{\mbox{\tiny F}}}$ is the Fermi velocity in the ferromagnets.
The usually small
spin dependence in the Fermi velocity has been neglected~\cite{GME1,GME2}.
The spin-dependent electrical current
is given by
\begin{equation}
{{\cal J}_s }\left( x,t \right) =  - {\sigma _s }\nabla {\mu _s }\left( x,t \right)\ ,
\label{Current}
\end{equation}
where ${\sigma _s } = {e^2}{\tau _s }k_{\mbox{\tiny F}}^3/6{\pi ^2}m$ is the Drude conductivity.

We consider first the drain electrode $B$. In this case, the
topological pure spin current ${\sigma _{\mbox{\tiny SH}}}E\left( t \right)$
flows from the QSH material
into the the electrode. The
boundary condition at the left edge $(x=0)$ of the electrode is given by
\begin{equation}
{{\cal J}_\uparrow }\left( 0,t \right) =  - {{\cal J}_\downarrow }\left( 0,t \right) = \frac{e}{\hbar }{\sigma _{\mbox{\tiny SH}}}E\left( t \right)\ ,
\label{Bound1}
\end{equation}
while that at the right edge $(x=L)$ reads
\begin{equation}
{\mu_ \uparrow }\left( L,t \right) = {\mu_ \downarrow }\left( L,t \right) = 0\ ,
\label{Bound2}
\end{equation}
since the right edge is grounded.
From Eqs.\ (\ref{DiffuseEq})-(\ref{Bound2}), one can readily obtain for
the spin-dependent electrochemical potential
\begin{equation}
{\mu _ \uparrow }\left( x,t \right) =  - {\mu _0}\frac{{\ell_{\mbox{\tiny D}}^2}}{{l_ \uparrow ^2}}{{e^{ - x/{\ell_{\mbox{\tiny D}}}}}}\ ,
\end{equation}
\begin{equation}
{\mu _ \downarrow }\left( x,t \right) = {\mu _0}\frac{{\ell_{\mbox{\tiny D}}^2}}{{l_ \downarrow ^2}}{{e^{ - x/{\ell_{\mbox{\tiny D}}}}}}\ ,
\end{equation}
with
\begin{equation*}
{\mu _0}\left( t \right) = \frac{e}{\hbar }{\sigma _{\mbox{\tiny SH}}}E\left( t \right){\ell_{\mbox{\tiny D}}}\left( {\frac{1}{{{\sigma _ \uparrow }}} + \frac{1}{{{\sigma _ \downarrow }}}} \right)\ ,
\end{equation*}
where $\ell_{\mbox{\tiny D}}^{ - 2} = {l_\uparrow^{ - 2} + l_\downarrow^{-2}}$ and $\ell_{\mbox{\tiny D}}$ is the spin-diffusion length. The electric voltage ${\mu}\left( x,t \right)$ in the ferromagnet is the average of spin up and down chemical potentials, ${\mu}\left( x,t \right) = \frac{1}{2}\left[ {{\mu _ \uparrow }\left( x,t \right) + {\mu _ \downarrow }\left( x,t \right)} \right]$.

The electric voltage in gate $A$ can be solved similarly. We find that the
electric voltage at the right edge of gate $A$ is equal to that at the left edge of gate $B$ in magnitude,
but with an opposite sign. As a result, the electric voltage difference between the
two inside edges of gates $A$ and $B$
is ${V_{\mbox{\tiny SH}}}\left( t \right) = 2{\mu}\left( 0,t \right)$, which can be derived to be
\begin{equation}
V_{\mbox{\tiny SH}}(t)=\frac{e}{\hbar }\sigma_{\mbox{\tiny SH}}E(t)\ell_{\mbox{\tiny D}}
\left(\frac{1}{{\sigma _ {\mbox{\tiny m}}}}-\frac{1}{{\sigma _ {\mbox{\tiny M}}}}\right).
\label{Voltage}
\end{equation}
Here, ${\sigma _ {\mbox{\tiny M}}}$ and ${\sigma _ {\mbox{\tiny m}}}$ are the majority-spin and minority-spin conductivities, as mentioned above. Therefore, by measuring
the electric voltage difference $V_{\mbox{\tiny SH}}(t)$,
the spin Hall conductivity $\sigma_{\mbox{\tiny SH}}$
 can be determined.

\section{Summary}\label{s5}

In summary, we have developed an analytical theory of the $ac$ QSH effect by
using the time-dependent scattering matrix method. The resulting $ac$ spin current flowing from the
QSH material into an electrode is linked to the winding numbers of the reflection matrix of the electrode, which also equal to the spin Chern numbers of the QSH system. The $ac$ QSH effect is a bulk transport phenomenon, being substantially different from its $dc$ counterpart, which relies on the existence of symmetry-protected gapless edge states.
A possible way to observe the $ac$ QSH effect experimentally was also suggested.

\begin{acknowledgments}
This work was supported by the State Key Program for Basic Researches of China under
grants numbers 2015CB921202 and 2014CB921103 (L.S.), the National Natural Science Foundation of China under grant numbers 11674160 and 11225420 (L.S.), and a project funded by the PAPD of Jiangsu Higher Education Institutions (L.S. and D.Y.X.). This work was also supported  by the U.S. Department of Energy, Office of
Basic Energy Sciences under Grant No. DE-FG02-06ER46305 (D.N.S).
\end{acknowledgments}


\begin{thebibliography}{99}
\bibitem{qshe1} C. L. Kane and E. J. Mele, Phys. Rev. Lett. \textbf{95}, 226801 (2005).

\bibitem{qshe2} B. A. Bernevig and S. C. Zhang, Phys. Rev. Lett. \textbf{96}, 106802 (2006).

\bibitem{qshe3} C. Wu, B. A. Bernevig, and S.-C. Zhang, Phys. Rev. Lett. {\bf 96}, 106401 (2006).

\bibitem{3dti1} J. E. Moore and L. Balents, Phys. Rev. B {\bf 75}, 121306(R) (2007).

\bibitem{3dti2} L. Fu and C. L. Kane, Phys. Rev. B {\bf 76}, 045302 (2007); L. Fu, C. L. Kane, and E. J. Mele, Phys. Rev. Lett. {\bf 98}, 106803 (2007).

\bibitem{3dti3} H. J. Zhang, C. X. Liu, X. L. Qi, X. Dai, Z, Fang, and S. C. Zhang, Nature Phys. {\bf 5}, 438 (2009).

\bibitem{rev1} M. Z. Hasan and C. L. Kane, Rev. Mod. Phys. \textbf{82}, 3045 (2010).

\bibitem{rev2} X. L. Qi and S. C. Zhang, Physics Today. \textbf{63}, 33 (2010).

\bibitem{rev3} Y. Ando, J. Phys. Soc. Japan {\bf 82}, 102001 (2013).

\bibitem{rev4} J. E. Moore, Nat. Phys. {\bf 11}, 897 (2015).

\bibitem{rev5} H. Weng, R. Yu, X. Hu, X. Dai, and Z. Fang, Adv. Phys. {\bf 64}, 227 (2015).

\bibitem{Weyl} B. A. Bernevig, Nat. Phys. {\bf 11}, 698 (2015).

\bibitem{rev6} Y. F. Ren, Z. H. Qiao, and Q. Niu, Rep. Prog. Phys. {\bf 79}, 066501 (2016).

\bibitem{Z2} C. L. Kane and E. J. Mele, Phys. Rev. Lett. {\bf 95}, 146802 (2005).

\bibitem{spinch1} D. N. Sheng, Z. Y. Weng, L. Sheng, and F. D. M. Haldane, Phys. Rev. Lett. {\bf 97}, 036808 (2006).

\bibitem{spinch2} E. Prodan, Phys. Rev. B \textbf{80}, 125327 (2009); E. Prodan, New J. Phys. \textbf{12}, 065003 (2010).

\bibitem{spinch3} H. C. Li, L. Sheng, D. N. Sheng, and D. Y. Xing, Phys. Rev. B \textbf{82}, 165104 (2010).

\bibitem{Yang} Y. Yang, Z. Xu, L. Sheng, B. G. Wang, D. Y. Xing, and D. N. Sheng, Phys. Rev. Lett. {\bf 107}, 066602 (2011).

\bibitem{HgTe} M. K\"{o}nig, S. Wiedmann, C. Br\"{u}ne, A. Roth, H. Buhmann, L. W. Molenkamp, X.-L. Qi, and S.-C. Zhang, Science {\bf 318}, 766 (2007).

\bibitem{RRDu} I. Knez and R.-R. Du, Frontiers of Phys. {\bf 7}, 200 (2012).

\bibitem{AcQSH} W. Y. Deng, H. Geng, W. Luo, W. Chen, L. Sheng, D. N. Sheng, and D. Y. Xing, arXiv:1606.08301 (2016).

\bibitem{Ac1} H. Jiao and G. E. W. Bauer, Phys. Rev. Lett. {\bf 110}, 217602 (2013).

\bibitem{Ac2} D. Wei, M. Obstbaum, M. Ribow, C. H. Back, and G. Woltersdorf, Nat. Commun. {\bf 5}, 3768 (2014).

\bibitem{Pump1} M. B\"{u}ttiker, H. Thomas, and A. Pr\^{e}tre, Z. Phys. B {\bf 94}, 133 (1994).

\bibitem{Pump2} P. W. Brouwer, Phys. Rev. B, {\bf 58}, R10135 (1998).

\bibitem{BHZ} B. A. Bernevig, T. L. Hughes, and S. C. Zhang, Science {\bf 314}, 1757 (2006).

\bibitem{InAs} C. X. Liu, T. L. Hughes, X. L. Qi, K. Wang, and S. C. Zhang, Phys. Rev. Lett. {\bf 100}, 236601 (2008).

\bibitem{HgTeSpinCh} H. Li, L. Sheng, R. Shen, L. B. Shao, B. Wang, D. N. Sheng, and D. Y. Xing, Phys. Rev. Lett. {\bf 110}, 266802 (2013).
    
\bibitem{Cpump3} M. N. Chen, L. Sheng, R. Shen, D. N. Sheng, and D. Y. Xing, Phys. Rev. B {\bf 91}, 125117 (2015).

\bibitem{Diffusion} S. Zhang, Phys. Rev. Lett. {\bf 85}, 393 (2000).

\bibitem{GME1} R. E. Camley and J. Barna\'{s}, Phys. Rev. B {\bf 664}, 266802 (1989).

\bibitem{GME2} M. Liu and D. Y. Xing, Phys. Rev. B {\bf 47}, 12272 (1993).


\end{thebibliography}
\end{document}